# CAOS Spectral Imager Design and Advanced High Dynamic Range FDMA-TDMA CAOS Mode

**MOHSIN A. MAZHAR [1] AND NABEEL A. RIZA [1,\*]**

[1] *Photonic Information Processing Systems Laboratory, School of Engineering, University College Cork, College Road, Cork, Ireland*
*\*Corresponding author:* n.riza@ucc.ie

**In the first part of the paper, a CAOS line camera is introduced for spectral imaging of one dimensional (1-D) or line targets. The proposed spectral camera uses both a diffraction grating as well as a cylindrical lens optics system to provide line imaging along the line pixels direction of the image axis and Fourier transforming operations in the orthogonal direction to provide line pixels optical spectrum analysis. The imager incorporates the Digital Micro-mirror Device (DMD)-based Coded Access Optical Sensor (CAOS) structure. The design includes a line-by-line scan option to enable two dimensional (2-D) spectral imaging. For the first time, demonstrated is line style spectral imaging using a 2850 K color temperature white light target illumination source along with visible band color bandpass filters and a moving mechanical pinhole to simulate a line target with individual pixels along 1-D that have unique spectral content. A ~412 nm to ~732 nm input target spectrum is measured using a 38 × 52 CAOS pixels spatial sampling grid providing a test image line of 38 pixels with each pixel providing a designed spectral resolution of ~6.2 nm. The spectral image is generated using the robust Code Division Multiple Access (CDMA) mode of the camera. The second part of the paper demonstrates for the first time the High Dynamic Range (HDR) operation of the Frequency Division Multiple Access (FDMA)-Time Division Multiple Access (TDMA) mode of the CAOS camera. The FDMA-TDMA mode also feature HDR recovery like the Frequency Modulation (FM)-TDMA mode, although at a much faster imaging rate and a higher Signal-to-Noise Ratio (SNR) as more than one CAOS pixel is extracted at a time. Experiments successfully demonstrate 66 dB HDR target recovery using both FDMA-TDMA and FM-TDMA modes, with the FDMA-TDMA mode operating at an encoding speed 8 times faster than the FM-TDMA mode given 8 FM channels are used for the FDMA-TDMA mode. The CAOS spectral imager can be used to image both full spectrum stationary line targets as well as spectrally map 2-D targets using line scanning methods. The demonstrated FDMA-TDMA CAOS mode is suited for improved speed and SNR linear HDR imaging.** © 2020 Optical Society of America

## 1. INTRODUCTION

It has been shown recently that extremely high (e.g., 96 dB) linear dynamic range spectrometry can be achieved using CAOS [1]. Specifically, the spectrometer demonstrated is a point spectrometer where the light emerging from a single point light source is spectrally resolved to generate the optical spectrum of the light source or point target sample under study for spectroscopy. The principle of point spectroscopy has been extended to line-based spectral imaging as well as line-by-line scan 2-D spectral imaging [2-3]. There are numerous applications that benefit from spectral imaging including remote sensing in agriculture [3-6] to artwork authentication [7-9] to food testing [10]. Numerous line and line-scan spectral imagers [11-19] have been designed and demonstrated using dispersive optics combined with both pixelated line sensors as well as pixelated 2-D image sensors including efforts to achieve high dynamic ranges (HDR) [20-21]. These instruments have so far been limited in linear dynamic ranges restricted by the 1-D line sensor and 2-D image sensor pixel performances that easily saturate and produce non-linear sensing responses such as commonly observed in the visible, Near IR bands using silicon CMOS and CCD sensors and also noted in eye safe > 1400 nm IR bands for FPA IR sensors. Such non-linearities lead to limited color contrast detection that negatively impacts spectral imaging, especially in remote sensing and medical imaging applications. Using the linear HDR CAOS point spectrometry results from ref.1, it is further possible to realize a CAOS line and line-scan spectral imager design that has the core capabilities of linear performances with high dynamic ranges to match spectral target HDR responses. Hence this paper first proposes and demonstrates the basic design and experimental operation of the basic proof-of-principle CAOS spectral line camera. The second part of the paper demonstrates for the first time the use of the advanced CAOS-mode called FDMA-TDMA that features the linear HDR image recovery powers of the FM-TDMA mode using Digital Signal Processing (DSP) based time-frequency domain spectrum analysis and electronic filtering, but at faster CAOS image capture rates and with improved SNR given all FDMA channels representing multiple CAOS pixels are simultaneously photo-detected. The paper includes both the design theory for the FDMA-TDMA mode as well as basic imaging experiments demonstrating the validity of the FDMA-TDMA mode.

## 2. CAOS LINE CAMERA FOR SPECTRAL IMAGING

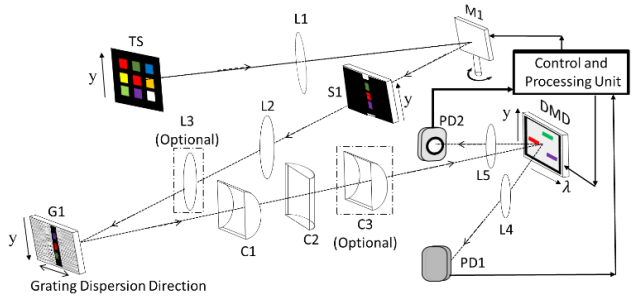

Figure 1. 3-D view of the linear HDR CAOS line scan spectral camera.

Fig.1 shows the 3-D view of a basic CAOS line camera design that functions as a linear HDR CAOS spectral imager to capture spectral content of both 1-D line and 2-D scenes. 2-D images are captured by two possible mechanisms. One approach shown in Fig.1 uses a line scan mirror M1 that rotates to scan the input image target scene TS across the fixed vertical slit S1 to allow time multiplexed projection of line-by-line input scene light into the CAOS spectral imager system. Alternately, the input scene can undergo physical 1-D translation across the slit when the platform between the camera and the scene is in relative motion such as when imaging objects are moving on a conveyer belt such as in factories. In general, L1 is the front lens imaging system that maps the TS plane image light onto the S1 plane. Assuming the special scenario later used for experiments in this paper where the light from the TS on the S1 plane is a collimated beam as the TS deployed is an optical fiber fed broadband light source and L1 is the fiber source collimation lens, L2 and L3 are used to form a telescopic image transfer system to present the collimated beam light image onto the grating optic G1 surface. Note that if the TS image light striking the S1 plane is not collimated, L2 and L3 are replaced by one spherical lens L2 that images the TS image light within the S1 slit plane onto the grating G1 plane. Hence L3 in Fig.1 is shown as an optional lens that depends on the spatial coherence properties of the light coming from the TS plane.

Next, the grating causes the optical spectrum of the image line pixels to spatially disperse along the grating frequency direction with the cylindrical lens C2 acting as a Fourier transforming lens along the spectrum spread direction. In effect, the spatially dispersed spectral components of each pixel in the imaged line fall along the DMD $\lambda$ labelled axis. Given FC2 is the focal length of C2, FC2 is the distance between G1 and C2 is and between C2 and the DMD. In contrast, given that the slit light on G1 is collimated (like in the experiment in this paper), the cylindrical lenses C1 and C3 form a slit illumination mapping system along the y-direction (i.e., slit direction) between the G1 and DMD planes thus preserving the slit mapping on the DMD [22]. Given these conditions, the inter-element distances are FC1 between G1 and C1, FC1 between C1 and C2, FC3 between C2 and C3, FC3 between C3 and DMD where C1 and C3 have focal lengths of FC1 and FC3, respectively. Given these constraints, CF2 = 2 CF1 and CF1=CF3. In effect, the spectrum of each pixel along the vertical line of the image shows up at its separate vertically positioned y location on the DMD for CAOS linear HDR imaging operations. Again, when the TS light falling on G1 is not collimated, only C1 is used to form an imaging system along the y-direction between G1 and the DMD making C3 an optional lens. Depending on the imaging demagnification/magnification required between G1 and the DMD planes along the y-direction, C1 can be placed either side of C2. Next, the DMD plane light illumination 2-D distribution is imaged using spherical imaging lenses L4 (with focal length F4) and L5 (with focal length F5) and point optical detectors PD1 and PD2, respectively. The PD1 and PD2 time-frequency CAOS signals are sent to the electronics for Radio Frequency (RF) style DSP for decoding operations and spectral line image extractions creating the desired linear HDR spectral imager via image line scans.

## 3. CAOS LINE CAMERA SPECTRAL IMAGING EXPERIMENTS

The Fig.1 CAOS line camera for spectral imaging is built in the laboratory (see Fig. 2). The illuminating test sample TS is the Avantes (UK) AvaLIGHT-HAL-S-Mini Pro-lite 2850 K bulb color temperature, 4.5 mW power, 360 – 2500 nm spectrum, 600 micron diameter fiber feed light source. L1 is a 5 cm focal length 2.5 cm diameter collimation lens with a 2.1 degree output beam divergence. In other words, the distance between TS and L1 is 5 cm to create a collimated light beam that strikes the S1 plane. A bandpass filter is placed in front of the collimating lens L1 for selective transmission of spectrum. The system also uses other components including: Vialux (Germany) DMD model V-7001 with micromirror size of 13.68 μm × 13.68 μm; DELL 5480 Latitude laptop for control and DSP, National Instruments 16-bit Analog-to-Digital Converter (ADC) model 6211; Thorlabs components that include bandpass filters FB700-40, FB650-10, FB620-10, FB600-40, FB550-10, FB450-40, FB450-10, and FB400-40, point PMT Model PMM02, 4 mm diameter Iris A1 (placed adjacent to the S1 plane), 5.08 cm diameter uncoated broadband lenses L2/L3 with focal lengths F2=5 cm, F3=6cm. Instead of a spherical lens L5, a spherical mirror SM1 of focal length of 3.81 cm is used in the system with one point PD labelled as PD2 that is the PMT. The cylindrical lenses C1/C2/C3 have focal lengths CF1=3 cm, CF2=6 cm, CF3=3 cm. G1 has a grating spatial frequency of $f_G$=600 lines/mm and $D_g$=1.62 nm/mrad dispersion at 750 nm. Key inter-component distances are: 11 cm between L1 and L2 giving an on-axis 6 mm diameter beam at the G1 plane with G1 rotated with incidence angle $\alpha$= 6 degrees. The distance are 21.2 cm between A1 and G1, 10 cm between DMD and L5/SM1 and 6.2 cm between L5/SM1 and point PD2/PMT.

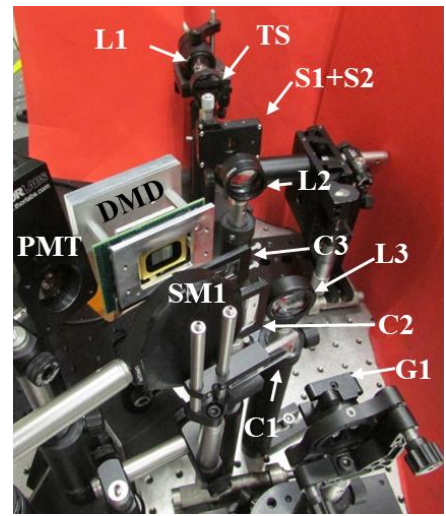

Figure 2. 3-D view of the experimental linear HDR CAOS line scan spectral camera.

To simulate a line in a scene for implementing spectral imaging, a vertically movable pixel in the vertical line of the scene is created using two crossed slits (S1 is vertical slit and S2 is horizontal slit) with micrometer adjustable apertures to control pixel size versus total optical power from pixel. The two crossed slits pixel motion along the slit S1 direction is also controlled with a micrometer stage so one can create multiple pixels in the line for spectral imaging. The white light fiber coupled source LS1 is used to illuminate the mechanical pixel in the vertical slit zone. The mechanical pixel in the vertical slit has an open air aperture and functions as white spectrum pixel in the line image creating a classic visible light spectrum on the DMD. This spectrum is then captured by the CAOS spectral imaging camera. As the white light pixel moves up and down using the mechanical stage motion, the captured wide spectrum of this white light pixel is also expected to move up and down in the captured spectral image. Indeed such behavior is observed for the test 500 micron × 500 micron size white light pixel that is mechanically moved along the slit axis.

observed by the CAOS spectral imager. Fig.4 shows captured spectral image data for pixels in a line that have used the following central wavelengths and 3-dB filter bandwidths: 700 nm (40 nm), 650 nm (10 nm), 620 nm (10 nm), 600 nm (40 nm), 550 nm (10 nm), and 450 nm (40 nm). Note that the CAOS image recovery for the 450 nm spectral filter is noisy using the CDMA mode given the 450 nm optical power levels for the LS1 source are low. For a better SNR recovery, the FM-TDMA [23] CAOS mode is used over a smaller DMD zone given the CDMA-mode has already located the spectral bin coarse location in the DMD space. Note that in the FM-TDMA mode, the CAOS pixels undergo FM time-frequency encoding using both the +θ and −θ tilt states of the DMD to produces a time varying unique frequency value square wave signal via the point photo-detector. This square wave signal is digitized and analyzed using the Fast Fourier Transform (FFT) algorithm to create the RF spectrum and recover the pixel irradiances.

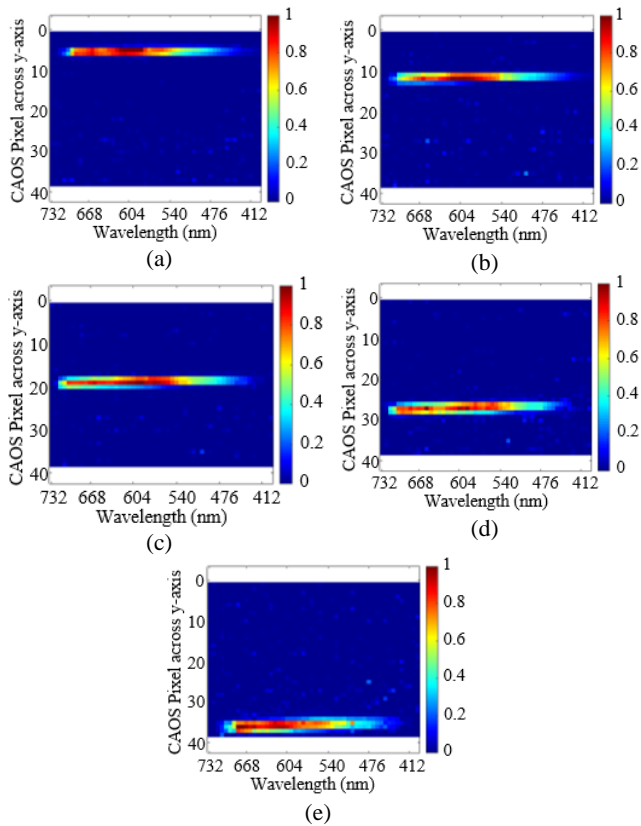

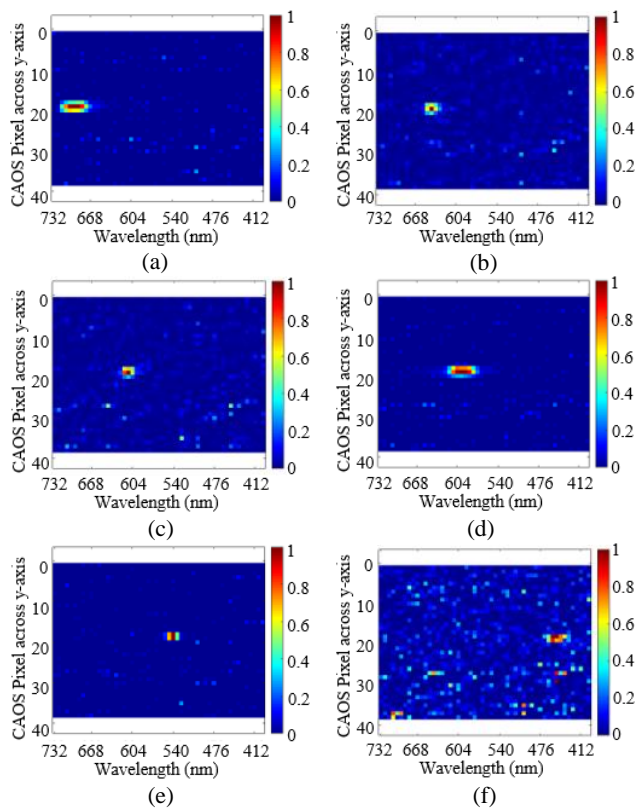

Figure 3. Line scan CAOS camera spectral images for a white light pixel moving along the vertical line direction of the incident spectral image. The x-axis is the wavelength axis with 52 CAOS pixels and the y-direction along the image vertical line with 38 CAOS pixels in the scene.

Fig.3 shows these experimental results confirming the operation of the designed CAOS line scan spectral camera as the white light pixel moves down along the y-direction of the scene line. These CAOS images are taken using the CAOS CDMA-mode [23] with a CAOS pixel size of 19×19 micro-mirrors, a CDMA Walsh code length of 2048 bits, a CDMA bit rate of 1 KHz, a DAQ sampling rate of 100 kSps to give a CDMA 2-D imaged 52 × 38 grid of 1976 CAOS pixels. The left side of the CAOS pixel x-axis (same as the λ-axis) covers a spectrum of ~732 nm to ~ 399 nm over the full width of the DMD that implements CAOS imaging. This spectral space is coarsely calibrated using fixed wavelength optical filters as also highlighted next. To demonstrate unique narrowband spectra of tested pixels in the vertical line of the scene, placed are select wavelength narrowband optical filters at the aperture of the white light mechanical pixel to create a unique narrow spectrum pixel that is

Figure 4. Line scan CAOS spectral camera images for narrow spectral band light pixel moving along its vertical line direction of an image. The x-axis is the wavelength axis corresponding to 52 CAOS pixels and the y-direction is along the image vertical line with 38 CAOS pixels in the scene. Test pixels spectral center wavelengths and 3-dB bandwidths used are: 700 nm (40 nm), 650 nm (10 nm), 620 nm (10 nm), 600 nm (40 nm), 550 nm (10 nm), and 450 nm (40 nm).

These improved images are shown in Fig.5 for the following spectral test pixel specifications: 450 nm (40 nm), 450 (10 nm) and 400 nm (40 nm). The FM-TDMA data is scaled by a factor of 24.66 to bring CDMA and FM-TDMA imaged data into the same scale for image observation. The FM-TDMA mode used a FFT processing gain of 48.16 dB, a sampling duration per CAOS pixel of 1 sec with 65535 samples per FM-TDMA time slot. Table 1 provides a summary of the measured CAOS mode test pixel SNR values showing why the CDMA mode had a dropping SNR value as the wavelengths of the test pixels got shorter as the spectrum has weaker light in the lower wavelength bands. Hence the FM-TDMA mode was used for spectral signal extraction as FM-TDMA engages FFT DSP gain via coherent signal processing.

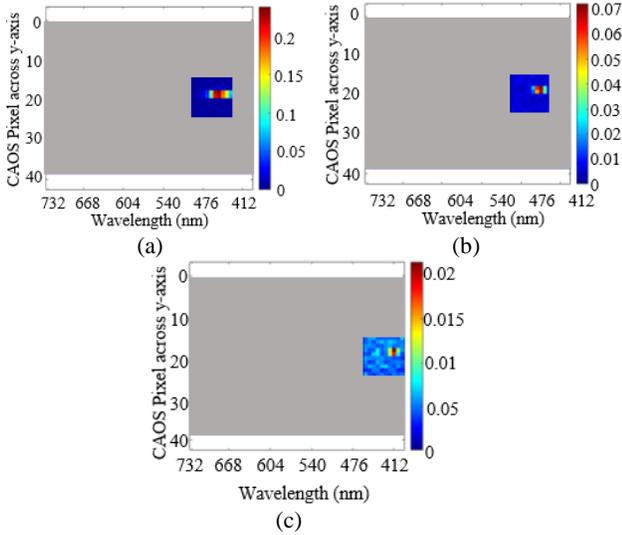

Figure 5. Line scan CAOS spectral camera images for narrow spectral band light pixel moving along the vertical line direction of the spectral image. The x-axis is the wavelength axis corresponding to 52 CAOS pixels and the y-direction is along the image vertical line with 38 CAOS pixels in the scene. Spectral test pixels used the following central wavelengths and 3-dB filter bandwidths: 450 nm (40 nm), 450 (10 nm) and 400 nm (40 nm).

Table 1. Line scan CAOS spectral camera test pixel wavelengths, 3-dB bandwidths (BWs), CAOS extraction modes, and measured minimum SNR for the imaged spectral pixels.

| Filter (nm) | Filter BW (nm) | CAOS Mode | Minimum SNR |
|---|---|---|---|
| 700 | 40 | CDMA | 69.9 |
| 650 | 10 | CDMA | 31 |
| 620 | 10 | CDMA | 23.7 |
| 600 | 40 | CDMA | 69.3 |
| 550 | 10 | CDMA | 87 |
| 450 | 40 | CDMA | 6.39 |
| 450 | 40 | FM-TDMA | 19.5 |
| 450 | 10 | FM-TDMA | 6.34 |
| 400 | 40 | FM-TDMA | 2.22 |

## 4. CAOS DIGITAL FDMA-TDMA MODE DESIGN THEORY

Expanding further via the CAOS FM-TDMA mode, this section presents the design details of the novel CAOS FDMA-TDMA mode [23] which maximizes the utilization of the full RF bandwidth available in the CAOS camera system. In comparison to FM-TDMA CAOS mode, the FDMA-TDMA CAOS-mode simultaneously deploys multiple frequency channels that are created and utilized for encoding and decoding of the chosen CAOS pixels. Specifically for the FDMA-TDMA mode, instead of a single CAOS pixel per frequency channel, multiple CAOS pixels undergo frequency modulation using different frequencies over a single TDMA time slot. Once the CAOS pixels irradiance encoded optical signal per TDMA slot is registered at the point PD and is digitized by the ADC, a different set of CAOS pixels are assigned the same RF carrier frequencies set and the encoding and CAOS pixels light irradiance extraction process is repeated. The source of digital FM behavior for this mode originates from the digital on-off FM modulation state of the TI DMD. Here, the frequency encoding of the CAOS pixels occurs when the DMD switches its micro-mirrors to realize on/off binary tilt angle states that result in time-varying photo-detected signal being a square wave with the fundamental frequency matching the designed CAOS pixel irradiance modulation carrier frequency. Note that the FDMA-TDMA mode is not limited to digital operation and an analog variant of FDMA-TDMA mode is also possible given the use of an analog operations Spatial Light Modulators (SLMs) and gray-scale modulation design DMD optical operations for line scan imaging [22].

Note that basics for several CAOS modes including the FDMA-TDMA mode are described previously in ref.23. This paper for the first time describes the detailed selection criterion for optimal channel frequency selection for the digital FDMA-TDMA CAOS mode via the TI DMD. The point PD registered and ADC digitized waveforms are a linear sum of the simultaneously harvested different CAOS pixel irradiances weighted square wave signals. Hence there is a certainty that the RF spectrum of the recorded point PD signal will include spectral content at frequency locations other than the designed carrier frequencies for the FDMA channels. For the FM-TDMA mode, such cross-channel RF spectrum noise is not possible given that there is no other FM source of spectral information in each TDMA slot other than the FM signal encoding the chosen single CAOS pixel, hence eliminating the possibility of inter-pixel crosstalk in the RF spectral decoding domain. This is one of the key reasons why best inter-pixel crosstalk linear HDR CAOS pixel extraction performance reaching 177 dB [24] is provided by the FM-TDMA mode, although this advantage comes at the cost of longer acquisition times as one CAOS pixel is encoded per TDMA time slot.

With this motivation, the FDMA-TDMA CAOS mode is aimed to alleviate this time constraint imposed by the FM-TDMA CAOS-mode while maintaining the HDR imaging capability. The use of simultaneous CAOS pixels encoded with different FM channels reduces the number of TDMA time slots by the a factor equal to the number of FM channels. However, for FDMA-TDMA the presence of RF spectral content from multiple CAOS pixels in a single TDMA time slot can lead to overlap and RF interference from different frequency carriers and their harmonics, resulting in increased inter-channel crosstalk which can severely degrade the reconstructed optical image via CAOS decoding. Therefore, a careful selection of FDMA-TDMA channel frequencies is critical to the robust and accurate operation of the FDMA-TDMA mode CAOS camera. In order to minimize crosstalk between the encoding FM RF digital carriers and their harmonics, an analysis of the spectral content of a single FM signal will aid in identifying additional FDMA channel frequencies. Let Eq.1 represent the ADC digitized periodic square wave for a CAOS pixel with FDMA-TDMA encoding where x[n] is given by:

$$x[n] = \begin{cases} 1, & |n| \leq N_1 \\ 0, & N_1 < |n| \leq N/2 \end{cases}$$

$$\text{and } x[n+N] = x[n] \quad (1)$$

where x[n] is the time-varying point PD detected digitized periodic square wave with a period N, and N1 denotes the number of samples for which x[n] is 1. Here n goes from –N/2, ….,-3,-2,-1,0, 1,2,3,….N/2 where N is even and $N_1$ is odd. Since the signal is periodic, the RF spectrum can be represented as a sum of Fourier coefficients [25]. The Fourier series coefficients for the signal described in Eqn.1 can be represented using the following expression:

$$a_k = \frac{1}{N} \sum_{n=-N_1}^{N_1} 1 \times e^{-jk\left(2\pi/N\right)n} \quad (2)$$

Here $k$ represents the Fourier series index. To simplify the Eqn.2, let $m = n + N_1$ and substitute it into Eqn.2 to write:

$$a_k = \frac{e^{jk\left(2\pi/N\right)N_1}}{N} \sum_{m=0}^{2N_1} e^{-jk\left(2\pi/N\right)m} \quad (3)$$

Evaluating the summation in Eqn. 3 gives:

$$a_k = \frac{e^{jk\left(2\pi/N\right)N_1}}{N} \left( \frac{1 - e^{-jk \times 2\pi(2N_1+1)/N}}{1 - e^{-jk \times 2\pi/N}} \right) \quad (4)$$

Eqn.4 can be further simplified as:

$$a_k = \frac{1}{N} \frac{e^{-jk\left(2\pi/N\right)}}{e^{-jk\left(2\pi/N\right)}} \left( \frac{e^{jk2\pi\left(N_1+\frac{1}{2}\right)N} - e^{-jk2\pi\left(N_1+\frac{1}{2}\right)/N}}{e^{jk\left(2\pi/2N\right)} - e^{-jk\left(2\pi/2N\right)}} \right) \quad (5)$$

The expression in Eqn.5 can be simplified further for the case when $k \neq 0, \pm N, 2N, \ldots$ to give:

$$a_k = \sin\left[ \frac{2\pi k \left(N_1 + \frac{1}{2}\right)}{N_1} \right] \times \frac{1}{N \times \sin\left(\pi k/N\right)} \quad (6)$$

Assuming that the square waveform has a classic duty cycle of 50%, Eqn.6 can be further simplified as $2N_1 + 1 = N/2$. Using this assumption, Eqn.6 can be rewritten as:

$$a_k = \frac{1}{N} \frac{\sin\left(\pi k/2\right)}{\sin\left(\pi k/N\right)} \quad (7)$$

It can be observed that for even values of $k$, the Fourier series coefficient is equal to zero as the numerator for the Eqn. 7 given by $\sin\left(\pi k/2\right)$ is zero for any even integer value of $k$. This result is key to selecting appropriate channel frequencies for the FDMA-TDMA mode as it gives critical knowledge of the frequency locations within the RF spectrum where the spectral contribution of the square wave is exactly zero. For odd harmonics of the carrier frequency, the Fourier series coefficient is significant and thus this frequency must be avoided for CAOS pixel irradiance encoding. The Fourier series coefficient for odd harmonics can be represented as:

$$a_k = \frac{1}{N \times \sin\left(\pi k/N\right)} \quad (8)$$

The conclusion that can be derived from the presented analysis is that the available frequency slots for FDMA RF channel selection with minimal crosstalk or minimal noise are even multiples of $k$, i.e., even harmonics of the fundamental carrier frequency of the encoding square wave. However, the analysis just mentioned does not take into account other FM signals present simultaneously, such as in the case of the CAOS FDMA-TDMA mode where multiple different FM signals are transmitting CAOS pixel irradiance encoded optical signals to the point PD in the CAOS camera. Specifically, the FDMA-TDMA mode would require that none of the harmonics from any of these FM encoding square waves for the different CAOS pixels overlap with any of the fundamental frequency components.

Tables 2, 3 and 4 show how the number of available frequency channels for FDMA encoding varies as more simultaneous FM channels are added to the CAOS camera system for the case when 4 FDMA channels are required. Let $f_a$ be the slowest and the first FM carrier selected for CAOS pixel irradiance encoding. Then as Table 2 shows, the available FDMA channel frequency slots are at FM carrier frequencies of $2f_a, 4f_a, 6f_a,$ and $8f_a$.

**Table 2. Shown are the available FDMA frequency channels when only one channel is utilized.**

| Frequency Channels Available | | | | |
|---|---|---|---|---|
| $f_1 = f_a$ | $2f_a$ | $4f_a$ | $6f_a$ | $8f_a$ |

For the introduction of a second FDMA frequency $f_2$, any of the available frequency channel slots can be chosen. Since the objective of the FDMA-TDMA mode is to minimize imaging time, a larger number of simultaneous CAOS pixels for FDMA encoding is desirable and for this objective, the closest channel frequency to $f_1$ must be assigned. With two FDMA frequencies, Table 2 would be modified and Table 3 shows the remaining available frequency channels. The × symbol represents the fact that this frequency is no longer available or is already being utilized. FDMA frequency $f_3$ must now be assigned $4f_a$ Hz which is an even harmonic of $f_1$, and $f_2$. As it is evident from Table 3, utilizing just another single-channel frequency has eliminated $6f_a$ as a possible frequency channel slot. The frequency $6f_a$ is indeed an even harmonic of $f_1 = f_a$, but this frequency is now an odd harmonic of $f_2$ i.e., $3f_2 = 6f_a$. Hence, this frequency channel will have a contribution from the third harmonic component $f_2$ which will act as noise. Similarly, for $f_4$ the closest available frequency channel slot $8f_a$ must be chosen.

**Table 3. Shown are the available FDMA frequency channels when a second channel frequency is selected.**

| Frequency Channels Available | | | | |
|---|---|---|---|---|
| $f_1 = f_a$ | $2f_a$ | $4f_a$ | $6f_a$ | $8f_a$ |
| × | $f_2 = 2f_a$ | $2f_2 = 4f_a$ | × | $4f_2 = 8f_a$ |

**Table 4. Shown are the selected FDMA frequency channels when four channel frequencies are selected.**

| Frequency Channels Available | | | | |
|---|---|---|---|---|
| $f_1 = f_a$ | $2f_a$ | $4f_a$ | $6f_a$ | $8f_a$ |
| × | $f_2 = 2f_a$ | $2f_2 = 4f_a$ | × | $4f_2 = 8f_a$ |
| × | × | $f_3 = 4f_a$ | × | $2f_3 = 4f_2 = 8f_a$ |
| × | × | × | × | $2f_3 = 4f_2 = 8f_a$ |

It is evident from an inspection of Table 3 that only frequencies that are geometric multiples of 2 are real candidates for the FDMA frequency channels. For the case just presented, $f_1 = f_a, f_2 = 2^1 f_a, f_3 = 2^2 f_a,$ and $f_4 = 2^3 f_a$. Equation 9.1 shows a fixed relationship between the first FDMA frequency and the next available FDMA frequency j-th channel slot for pixel irradiance encoding with minimal crosstalk and highest SNR performance to be given by:

$$f_j = 2^{j-1} f_1, \quad (9.1)$$

$$f_1 = 2^{m-1} \Delta f, \quad (9.2)$$

where j=1,2,3......,P and $f_j$ represents the j-th FDMA channel frequency. $f_1 = f_a$ is a user chosen lowest FDMA channel frequency that can be given by Eqn.9.2 that is available for use where m=1, 2, 3.... and Δf=1/T where T is the TDMA time slot duration in seconds representing the CAOS pixel FDMA encoding window. Both expressions in Eqn.9 without loss of generality satisfy the criteria illustrated via Tables 2 to 4 that any encoding FM frequency for the FDMA-TDMA mode satisfies a power of 2 factor relationship with the fundamental frequency carrier. Eqn.9.2 also illustrates that full, i.e., complete FM square wave cycles are enclosed in the time window T, e.g., m=1 implies that the lowest FDMA frequency $f_1$ has 1 complete cycle within T or m=3 indicates an FDMA frequency $f_1$ with 4 complete cycles with the T seconds window. Fig.6(a) in the next section indeed visually illustrates this FDMA encoding signal design that also prevents capture of partial square wave cycles that can lead to unwanted spectral decoding noise. In effect, any of the FDMA signals shown in Fig.6(a) can be used as the lowest FDMA channel $f_1$ frequency.

Thus Eqn.9 can be used to design all the required digital FDMA-TDMA CAOS-mode FM channel carrier frequencies providing minimal inter-pixel crosstalk as there would be minimal inter-frequency channel crosstalk between the selected FDMA encoding channels. Furthermore, the ADC sampling rate $f_s$ (Sps) is chosen such that a power of 2 number of signal samples Q is generated to prevent the FFT DSP algorithm from adding any zero value samples in the data set to ensure that Q is a power of 2 as required by the FFT radix-2 algorithm [26]. Hence, the CAOS system is designed for FM modes so that Q= $f_s$ T= $2^p$ where p is an integer. One can also write $f_s = 2^p \Delta f$ indicating how the ADC sampling frequency should be selected. The FFT spectral resolution is also given by Δf (Hz) = $f_s$/Q= 1/T. Following Eqn.9 for the FDMA channels frequencies selection as well the mentioned condition of the $f_1$ frequency and requirements for $f_s$, T, and Q , one then ensures that each FDMA channel falls wholly within a unique spectral bin within the full FFT spectrum of multiple individual spectral bins. For example, if T=0.25 seconds, Δf=4 Hz and the spectral bins occur at 0, Δf, 2Δf, 3Δf,..... Hz locations of the FFT spectrum. Thus one prevents inter-pixel crosstalk as no spectral content of an FDMA frequency channel partially overlaps another spectral bin containing another FDMA frequency channel. Also using Eqn.9.2 with m=5 and Δf = 4 Hz, $f_1 = f_a$ = 64 Hz. Note that all the mentioned design criteria must be maintained when adjusting the RF spectral FFT DSP gain.

## 5. CAOS DIGITAL FDMA-MODE DESIGN SIMULATION

In order to verify the analysis presented in the previous section, the CAOS digital FDMA-TDMA mode was modeled in MATLAB. Eight CAOS pixels with the same irradiance levels were imaged using the FDMA-TDMA mode using just one TDMA time slot. The FDMA frequency selection criterion described previously was adhered. Pixel irradiance encoding FDMA channel frequencies used were $f_a$ = 64 Hz, $f_j$ = $2^{j-1} f_a$, where the index j goes from 1 to 8, implying that the 8 FDMA channel frequencies chosen are $f_1$=64, $f_2$=128, $f_3$=256, $f_4$=512, $f_5$=1024, $f_6$=2048, $f_7$ = 4096 Hz and $f_8$= 8192 Hz. Here, T=1 sec so Δf=1 Hz and m=7 giving $f_1=f_a$=64 Hz.

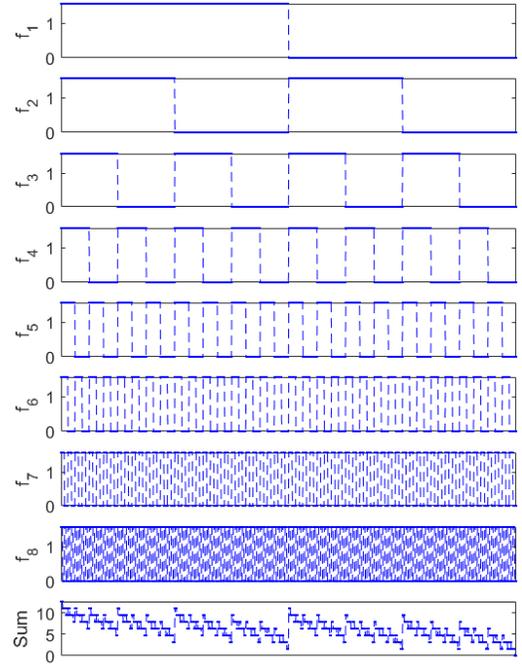

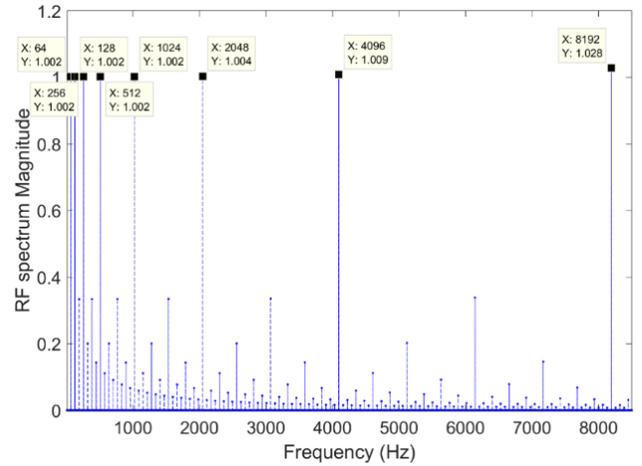

Figure 6. (a) Shown is the time domain signal of all encoded pixel irradiances and their photo-detected sum. The y-axis represents the amplitude of the shown nine signals. The time domain signals are only shown for a short time interval of 1/(64 Hz) = 0.0156 seconds. Figure 6. (b). Shown is the RF spectrum magnitudes recorded using the entire TDMA time slot photo-detected data. Equal magnitude RF spectral peaks can be seen at 64,128,256,512,1024,2048,4096, and 8192 Hz, which matches the designed pixel irradiances. Note harmonics present are overlapping at non-FDMA frequency channel slots.

The recorded time-varying waveform resulting from all the FDMA encoded CAOS pixels was spectrally processed with the FFT radix-2 algorithm for RF spectrum analysis. Figure 6 (a) shows the encoded pixel irradiances for all the FDMA-mode carrier frequencies where the y-axis marking $f_j$ shows the frequency used to encode the pixel irradiances. The plots shown are for a total time duration of 1/(64 Hz) = 0.0156 seconds, which is the period of one complete $f_1$ =64 Hz cycle. The last plot in Figure 6(a) shows a time domain weighted sum signal that is formed by adding the eight encoded pixel irradiance signals shown above it. Fig.6 (b) shows the RF spectrum observed for this cumulative photo-detected signal. Clear distinct RF spectral carrier peaks show up with equal strength at $f_1$= 64 Hz, $f_2$=128 Hz, $f_3$= 256 Hz, $f_4$=512 Hz, $f_5$=1024 Hz, $f_6$ = 2048 Hz, $f_7$ = 4096 Hz and $f_8$ = 8192 Hz, confirming no spectral overlap or inter-pixel interference.

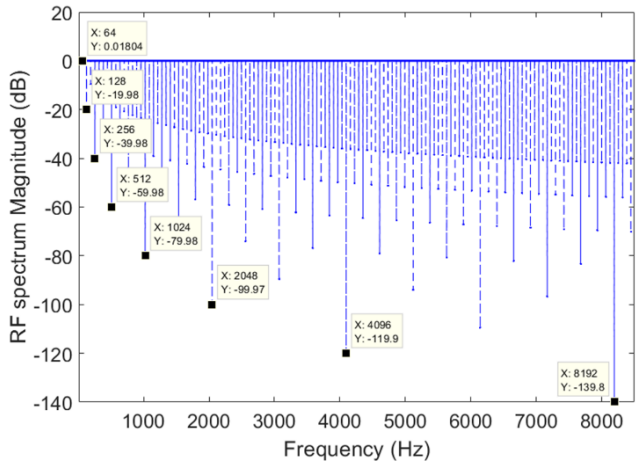

Figure 7. Shown is accurate pixel irradiance recovery for all eight FDMA encoded pixels with a maximum dynamic range of 140 dB. RF spectrum magnitude recovered at 64,128,256,512,1024,2048,4096, and 8192 Hz match the designed pixel irradiances and the designed dynamic range.

**Table 5. Presented for the simulation described is a comparison of the designed pixel irradiances and dynamic range against the recovered pixel irradiances and recovered dynamic range.**

| FDMA Frequency Channel (Hz) | Designed Pixel Irradiance (Normalized) | Recovered Pixel Irradiance (Normalized) | Designed Dynamic Range (dB) | Recovered Dynamic Range (dB) |
| --- | --- | --- | --- | --- |
| 64 | 1 | 1 | - | - |
| 128 | $1 \times 10^{-1}$ | $1.002 \times 10^{-1}$ | 20 | 19.99 |
| 256 | $1 \times 10^{-2}$ | $1.002 \times 10^{-2}$ | 40 | 39.99 |
| 512 | $1 \times 10^{-3}$ | $1.002 \times 10^{-3}$ | 60 | 59.99 |
| 1024 | $1 \times 10^{-4}$ | $1.002 \times 10^{-4}$ | 80 | 79.99 |
| 2048 | $1 \times 10^{-5}$ | $1.004 \times 10^{-5}$ | 100 | 99.99 |
| 4096 | $1 \times 10^{-6}$ | $1.009 \times 10^{-6}$ | 120 | 119.94 |
| 8192 | $1 \times 10^{-7}$ | $1.028 \times 10^{-7}$ | 140 | 139.78 |

Next to test the HDR capabilities of the CAOS FDMA-TDMA mode, another simulation was conducted. This time each of the eight CAOS pixels was allocated a different irradiance where the dynamic range between the brightest and weakest pixels was set to a HDR value of 140 dB (or $10^7$ to 1 in relative irradiance levels, i.e., DynamicRange (DR)=20×log($10^7$/1)= 140 dB). Figure 7 shows the RF spectrum magnitude in the DR dB scale observed for each of the chosen FDMA-TDMA frequencies of $f_1$ =64 Hz, $f_2$ =128 Hz, $f_3$ =256 Hz, $f_4$ =512 Hz, $f_5$ =1024 Hz, $f_6$ =2048 Hz, $f_7$ = 4096 Hz and $f_8$ = 8192 Hz. The CAOS pixel with the brightest pixel irradiance was encoded with a frequency of $f_1$ = 64 Hz, whereas the FDMA frequency channel allocated to the weakest irradiance pixel was $f_8$ = 8192 Hz. The irradiances recovered yield a computed HDR= 139.8 dB in comparison to the designed 140 dB. Here, the HDR is computed using HDR (dB) = 20×log($I_{max}/I_{min}$), where $I_{max}$ and $I_{min}$ are the recorded maximum and minimum pixel irradiances, respectively. The simulation results confirm the minimal crosstalk and robust performance of the FDMA-TDMA mode of the CAOS camera. Table 5 shows designed input pixel irradiance and dynamic range values along with the recovered irradiances and dynamic range values. Note that both basic simulations were done without the presence of any external noise sources, i.e., electronic noise or white noise.

## 6. CAOS DIGITAL FDMA-MODE IMAGING EXPERIMENT

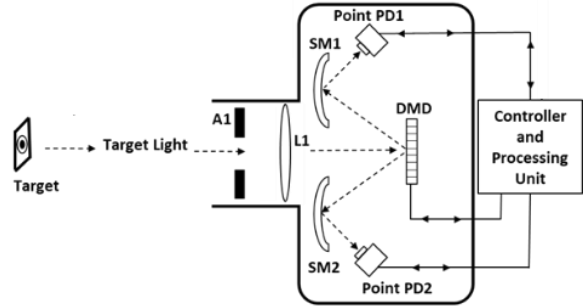

Figure 8. Top view of the CAOS camera design used to demonstrate the CAOS FDMA-TDMA mode

Figure 8 shows a design of the demonstrated FDMA-TDMA mode CAOS camera. The light transmitted travels to the main imaging lens L1 through aperture A1 and forms a target image onto the DMD. Depending on the mode engaged for the CAOS camera, the DMD device encodes pixel irradiances. As explained earlier, for the case of FM-TDMA all the CAOS pixels are sequentially imaged with the same FM carrier frequency which is generated by switching the micro-mirrors of the DMD to an on/off state at a frame rate twice that of the FM carrier frequency. In contrast for FDMA-TDMA, multiple CAOS pixels are encoded in the same TDMA slot using different carrier frequencies and the process is repeated sequentially. For both FM-TDMA and FDMA-TDMA, the scene light is imaged onto the point PD1 and point PD2 by the spherical mirrors SM1 and SM2, respectively. These photo-detected light signals are digitized with the aid of an ADC and then processed using DSP correlation and spectral analysis techniques. The CAOS camera design shown in Fig.8 using the PD1 only channel was built in the laboratory with the following equipment: Vialux DMD model V-70001, 16-bit National Instruments ADC model USB-6366, and DELL 5480 laptop as control and processing unit. Optical components from Thorlabs include a variable diameter iris A1, a 5.08 cm diameter 10 cm focal length lens L1, 5.08 cm diameter spherical mirror SM1 with a focal length of 3.81 cm, and PDA100A2 as point PD1. Inter-component distances are: A1: L1 4 cm, L1:DMD 11.6 cm, DMD:SM1 10 cm, SM1: PD1 6.2 cm.

To begin imaging, a 5 W flashlight is used as a test target. The flashlight is placed at a distance of 71 cm from L1 and the input light from the flashlight forms an image at the DMD plane. First, the CDMA mode of the CAOS camera is engaged for imaging using a total of 3600 CAOS pixels with a grid configuration of 60×60 CAOS pixels with each CAOS pixel of size 6×6 micro-mirrors and a DMD-based CDMA code bit rate of 1 kHz. A Walsh code length of 4096 is used to image this grid. Figure 9 (a) shows the results produced for this mode. Next, the FM-TDMA mode of the CAOS camera is engaged. Here, a FM carrier frequency of 8192 Hz is chosen to encode pixel irradiances using a DMD frame rate of 16384 Hz. A total of 2048 on/off cycles are captured using the point PD1 giving a pixel TDMA slot dwell time T of 0.25 seconds giving a $\Delta f$= 4 Hz. An ADC sampling rate of $f_s$= $2^p \Delta f$ =65536 Sps for p=12 is chosen giving Q=16384 digitized samples per TDMA slot. These 16384 samples are processed using the FFT radix-2 algorithm offering a 10×log(16384/2) = 36.12 dB processing gain [26]. The image acquisition time for the FM-TDMA mode was 3600×0.25 = 900 seconds. Figure 9 (b) shows the image captured using the FM-TDMA mode. Next, the FDMA-TDMA mode of the CAOS camera was engaged. In order to establish a performance comparison between FM-TDMA and FDMA-TDMA modes, all the acquisition parameters must be the same, i.e., acquisition time per TDMA slot, DSP FFT gain and ADC sampling rate. Using $f_1$= 128 Hz as the lowest FDMA channel frequency and the relationship described in Eqn.9, $f_2$= 256 Hz, $f_3$= 512 Hz, $f_4$= 1024 Hz, $f_5$= 2048 Hz, $f_6$= 4096 Hz, and $f_7$= 8192 Hz were selected.

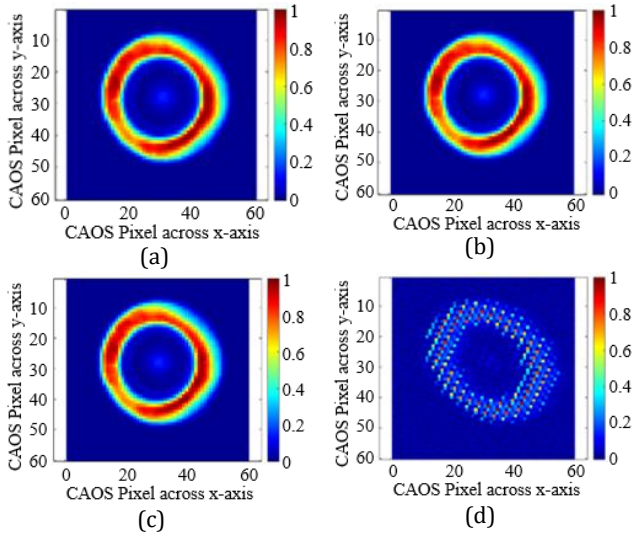

Figure 9. Shown are the images recorded by the CAOS camera using different imaging modes. (a) Image captured using CDMA mode, (b) Image captured using FM-TDMA and (c) Image captured using the FDMA-TDMA mode following the criteria defined in Eqn.9. (d) Shown is the CAOS image captured using the FDMA-TDMA mode for which Eqn.9 is not fully satisfied.

Figure 9 (c) shows the image captured using the FDMA-TDMA mode. The image acquired took a total of [3600/7]×0.25=128.75 seconds; here the symbol with half brackets ⌈ ⌉ represents the ceiling function. This function operates as follows: 3600/7 = 514.29, ⌈3600/7⌉ = 515. Clearly, the FDMA-TDMA mode takes a shorter time to image the same target. For comparison using the FDMA-TDMA mode, the same target was imaged 900/128.75 = 6.99 times faster than the FM-TDMA mode indicating a significant 7 times improvement in faster imaging speed.

Note that the CAOS FM-based signal encoding and DSP decoding methods are designed to meet the constraints mentioned in section 4, including those for $f_1$, $f_s$, T, and Q. The lowest frequency $f_1$ =128 Hz and is per design kept higher than the 50 Hz AC power mains fundamental frequency to avoid spectral crosstalk effects. In addition, $f_1$ is kept high enough to minimize 1/f electronic noise contributions in the decoding DSP process. Given ∆f=4 Hz, 128 Hz falls in the 22nd spectral bin of the FFT spectrum. Another choice for $f_1$ can be 64 Hz as it falls in the 18th spectral bin and is higher than 50 Hz. With T=0.25 s for ∆f= 4 Hz, note that both 64 Hz and 128 Hz satisfy the condition $f_1=\Delta f\, 2^m$.

To further test the validity of the proposed FDMA-TDMA mode and its design theory, 4 out of the previously determined FDMA-TDMA frequencies are changed so that these 4 frequencies do not satisfy the criterion defined in Eqn.9. The 7 channel frequencies now are $f_1$= 1170.3 Hz, $f_2$=1368.3 Hz, $f_3$= 1638.4 Hz, $f_4$= 2048 Hz, $f_5$= 2730.6 Hz, $f_6$= 4096 Hz, and $f_7$ = 8192 Hz. Amongst these FDMA channel frequencies, $f_1$, $f_2$, $f_3$, and $f_5$ do not obey the criterion described in Eqn.9. Figure 9 (d) shows the CAOS image recorded using the FDMA-TDMA mode with some of the channel frequencies non-optimal as mentioned. The image captured using these new set of frequencies is severely distorted and has unwanted artifacts thereby showcasing the importance of correct frequency selection for all FDMA channels for robust imaging using the demonstrated digital FDMA-TDMA CAOS mode.

To test the proposed FDMA-TDMA mode for a HDR scenario, the Image Engineering (Germany) lightbox LG3 is placed at a distance of 129 cm from L1 and the distance between L1 and DMD is adjusted to a distance of 10.8 cm to satisfy the imaging condition. A custom-made 66 dB transmissive HDR target is placed at the illuminating surface plane of LG3. This custom target was made using Thorlabs absorptive attenuating Neutral Density (ND) filters. Figure 10 presents the target consisting of 6 circular patches arranged in a 2×3 configuration. Since the target being imaged is an HDR target, the DSP FFT gain for both FM-TDMA and FDMA-TDMA modes is increased to 10×log(65536/2) = 45.16 dB and the ADC sampling rate is fixed at $f_s$= $2^p$ ∆f = 65536 Sps where p=16. This additional processing gain comes at the expense of a longer acquisition time. Point PD data is now acquired for a total of T=1 second for each TDMA time slot giving a ∆f=1 Hz. In order to minimize the total acquisition time an additional FDMA frequency channel is introduced using an $f_a$= 64 Hz, meaning the eight FDMA channel frequencies now are $f_1$= 64 Hz, $f_2$ = 128 Hz, $f_3$= 256 Hz, $f_4$= 512 Hz, $f_5$= 1024 Hz, $f_6$= 2048 Hz, $f_7$= 4096 Hz, and $f_8$= 8192 Hz.

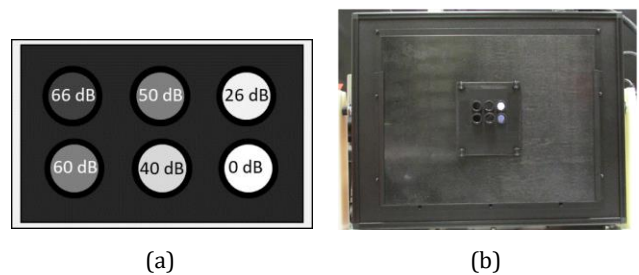

Figure 10. (a) Shown is the custom-made 66-dB HDR transmissive test target with attenuation levels set at 0 dB, 26 dB, 40 dB, 50 dB, 60 dB, and 66 dB. (b) Shown is the target mounted on the LG3 lightbox.

To begin imaging a 44×29 CAOS pixel-imaging grid with each CAOS pixel of size 8×8 micro-mirrors is identified. Figure 11(a) and 11 (b) show the target image captured using FM-TDMA and FDMA-TDMA modes, respectively. The target is recovered completely by both CAOS modes. Figure 12 shows pixel dynamic range plots for two horizontal lines of CAOS pixels passing through the target patches. These two horizontal lines are chosen such that one passes through the center of the patches in the top half of the 66-dB target whereas the other passes through the center of the patches in the bottom half of the target. Figure 12 (a) and (b) are plots for pixel dynamic ranges for the top and bottom half patches recovered using FM-TDMA mode, respectively. Similarly, Fig.12 (c) and (d) present pixel dynamic range plots of the top half and the bottom half of the target recovered using FDMA-TDMA, respectively. Flat tops within each image scan line plot indicate the detection of a uniform irradiance level of each patch within the target grid.

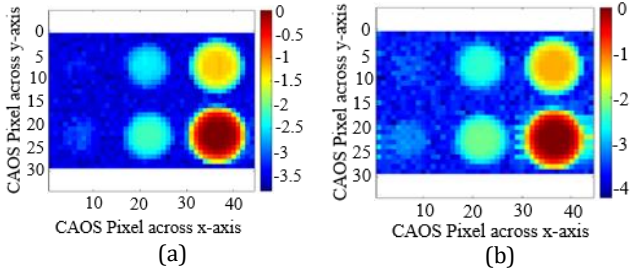

Figure 11. Shown are the HDR images captured using the (a) FM-TDMA and (b) FDMA-TDMA modes of the CAOS camera. Images shown are plotted on a log scale for ease of viewing.

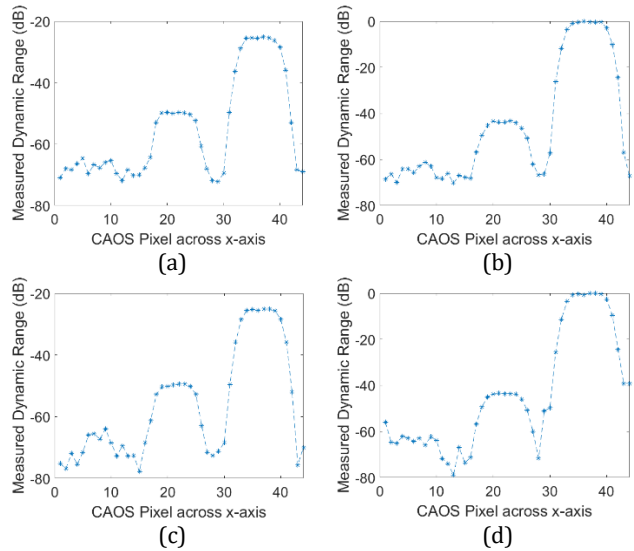

Figure 12. Shown are the measured pixel dynamic ranges across the horizontal axis at two different heights of the CAOS image. (a), (b) show FM-TDMA measured pixel dynamic ranges for the top and bottom half of the target whereas (c), (d) show pixel dynamic ranges measured using FDMA-TDMA mode at the same image scan line locations.

**Table 6. FM-TDMA mode CAOS Camera designed and measured target patch Dynamic Range (DR) and minimum SNR values.**

| Norm. Irradiance (Designed) | Norm. Irradiance (Measured) | Designed DR (dB) | Measured DR (dB) | Minimum SNR |
|---|---|---|---|---|
| 1 | 1 | 0 | 0 | 2547 |
| 0.05 | 0.054 | 26 | 25.3 | 138.2 |
| 0.01 | 0.0065 | 40 | 43.8 | 16.5 |
| 0.0032 | 0.0032 | 50 | 49.9 | 8.1 |
| 0.001 | 0.0007 | 60 | 62.8 | 1.8 |
| 0.00032 | 0.0005 | 66 | 66.8 | 1.2 |

**Table 7. FDMA-TDMA mode CAOS Camera designed and measured target patch DR and minimum SNR values.**

| Norm. Irradiance (Designed) | Norm. Irradiance (Measured) | Designed DR (dB) | Measured DR (dB) | Minimum SNR |
|---|---|---|---|---|
| 1 | 1 | 0 | 0 | 3169 |
| 0.05 | 0.054 | 26 | 25.3 | 172 |
| 0.01 | 0.0065 | 40 | 43.8 | 20.5 |
| 0.0032 | 0.0032 | 50 | 49.9 | 10.1 |
| 0.001 | 0.0007 | 60 | 62.8 | 2.3 |
| 0.0005 | 0.0004 | 66 | 66.95 | 1.4 |

Table 6 and Table 7 present a comparison of the implemented FM-TDMA and FDMA-TDMA modes against the designed dynamic ranges of the 6 patch target. It is evident that the target has been fully recovered by both FM-TDMA and FDMA-TDMA modes. Another indicator of robust recovery is the CAOS extraction measurement minimum SNR>1 value for each of the target patches. FM-TDMA mode recovers the target dynamic range of 66-dB within 1 dB of the design with an SNR of 1.2 whereas FDMA-TDMA mode matches this result and recovers the target within 1 dB dynamic range difference while giving a slightly higher SNR value of 1.4. The image encoding time for the FM-TDMA mode for the CAOS grid of 1276 pixels is 1276×1 seconds = 1276 seconds. In contrast the FDMA-TDMA mode encoding time for the same CAOS grid is ⌈1276/8⌉×1=160 seconds; again here ⌈ ⌉ represents the ceiling function. This FDMA-TDMA mode speed advantage of 7.98 or near 8 times versus the FM-TDMA mode coupled with the evidence of SNR robust HDR recovery showcases the power of the proposed FDMA-TDMA mode. Here, given the extreme speed of DSP processors for CAOS decoding operations, it is the encoding operation that dominates the total imaging time and hence the focus of the analysis for the two CAOS HDR capability modes. Note that the method adopted to measure the imager noise relies on taking the average of irradiance values of the black area pixels corresponding to no light regions in the imaged target scene. The signal value is computing by taking the average of the target specific patch measured pixel irradiances. The 10 log of the ratio of the signal versus noise values gives the SNR in dB and a detailed explanation can also be found in ref 24.

## 7. CONCLUSION

For the first time, designed and demonstrated is a proof-of-principle CAOS spectral camera. The design engages a line scan mechanism to enable 2-D spectral imaging while a no-scan starring-mode design allows line-type CAOS spectral imaging. Experiments are conducted using a moving test pixel over a vertical slit to simulate a line scene. Various spectral content for the test pixel are deployed such as white broad visible spectrum light to 10 nm narrow bandwidth single wavelength pixels. CDMA and FM-TDMA CAOS modes are deployed to successfully image these test pixels, demonstrating the feasibility of the proposed CAOS spectral camera. Future work involves accurate spectral calibration of the imager as well as optical design optimizations to enable a more efficient system for better SNR performances. In the second part of the paper, a novel CAOS FDMA-TDMA mode has been designed and demonstrated. Specifically, the digital FDMA-TDMA mode is proposed that aims at maximizing the use of RF bandwidth available in the CAOS system while maximizing the speed of imaging operation. In comparison to the FM-TDMA CAOS mode, the FDMA-TDMA mode offers a much higher imaging speed and the same HDR imaging capabilities with minimal crosstalk. A key factor for the robust operation of the digital FDMA-TDMA mode is the optimal FDMA channels frequencies selection for the digital SLM. Hence a digital FDMA frequency selection design criterion is described in detail. Experiments conducted with a 66-dB HDR target show 8-channel digital FDMA-TDMA HDR target recovery matches FM-TDMA mode recovery and uses a much shorter factor of 8 reduced CAOS pixels encoding time. The demonstrated two imaging innovations can impact various applications including industrial metrology and inspection and biomedical imaging.

**Disclosures**. The authors declare no conflict of interest.